# MASS-DIAMETER RELATION OF GLOBULAR STAR CLUSTERS, ELLIPTICAL GALAXIES AND SPHERICAL CLUSTERS OF GALAXIES

JOHANN ALBERS

Fachbereich Physik der Universität des Saarlandes

66041 Saarbrücken, Germany

## ABSTRACT

In a preceeding paper alternative reflections on gravitation were developed. There it was assumed that the primary interaction between two masses is not of attractive but of repulsive nature. The repulsive force results from the impuls transfer produced by the gravitational radiation which is emitted and absorbed by both masses. The observed attractive force between the two masses according to Newton`s law of gravitation, however, is a secondary effect and a consequence of the existence of all the masses in the universe. The mutual screening of the gravitational radiation of all masses of the universe by the two masses under consideration leads to the gravitational attraction between them. The balance between primary, repulsive and secondary, attractive forces can stabilize highly concentrated spherical mass accumulations with a linear dependence of their mass on the square of their diameter. Such objects can really be observed in the universe in the form of globular star clusters, elliptical galaxies and spherical clusters of galaxies. The scatter of the data of every group is rather large. But the collection of the objects of all three groups, reaching from the smallest globular star cluster to the largest spherical cluster of galaxies, with masses differing by almost 12 orders of magnitude, clearly shows the proposed mass-diameter relation.

*Subject headings:* gravitation - globular clusters - elliptical galaxies - galaxies: spherical clusters - masses - diameters

## 1. INTRODUCTION

The main points of the alternative reflections developed in the foregoing paper (Albers 1997) are recapitulated here shortly to an extent necessary for the understanding of the special points discussed in more detail in this paper: Every mass emits and absorbs gravitational radiation which is treated in complete analogy to the laws of optics and electromagnetic radiation. At a certain reference point Q the intensity I of the gravitational radiation ( GR ) is achieved by summing up the GR of all masses $m_i$ surrounding Q at distance $R_i$

$$I = \text{Sum}_i (E * m_i / R_i^2). \qquad (1)$$

E is a factor of proportionality that describes the efficiency to emit gravitational radiation per unit of mass. E is taken in the following as a constant, the simplest assumption which is probably not valid independent of the mass concentration with values between the mean mass density of the universe of about $5*10^{-28}$ kg/m$^3$ (Unsöld & Baschek 1991) and the density of stars of about $10^3$ kg/m$^3$. The emissivity may also depend on other parameters like temperature etc. If, however, only objects in the universe are considered which are comparable with regard to the relevant parameters, then the parameter-dependent emissivity may be replaced by a constant E. The quantity I in equation (1) is named intensity, although it is not identical but only proportional to the physical intensity defined and measured as energy flux per second and per unit area.

If the effect of all masses $m_j$ of the universe is summed up according to equation (1) for a reference point Q one gets a certain intensity Io

$$I_o = \text{Sum}_j (E * m_j / R_j^2). \qquad (2)$$

This intensity Io, however, is dependent on the reference point Q and will be very high if Q is located at the center of a spherical cluster of galaxies and will be low if Q is far away from any mass accumulation, as explained in the preceeding paper (Albers 1997). From this intensity $I_o$ there results an attractive force F between every two masses $m_1$ and $m_2$ located near Q

$$F = - I_o * A*m_1 * A*m_2 * P / r^2. \qquad (3)$$

Here $A*m_1$ represents the absorption caused by the mass $m_1$ of the gravitational radiation with intensity $I_o$ directed towards $m_2$. From this absorbed part $I_o*A*m_1$ the absorption by $m_2$ is missing which leads to the factor $A*m_2$. The factor P represents the proportionality between the absorbed radiation and the impuls transfer by the absorption of thís radiation. (Here the term absorption is representative for all similar processes which can lead to such an impuls transfer).

The proportionality of F with $r^{-2}$, where r is the distance between $m_1$ and $m_2$, with the exact power of 2, is a consequence of the application of the laws of optics on these phenomena.

Equation (3) is equivalent to Newton´s law of gravitation

$$F` = - G * m_1 * m_2 / r^2. \qquad (4)$$

The „universal constant G of gravitation" of this law, however, which is equal to $I_o*A*A*P$ in equation (3) is no constant but a factor (gravitational factor) depending on the mass-distribution in the universe and the position of the reference point Q. For a first rough estimate of the attractive force between two masses $m_1$ and $m_2$ according to equation (3) it may be assumed in the following that the factor $I_o$, described in equation (2), is a constant.

The attractive interaction between the two masses $m_1$ and $m_2$ according to equation (3) is a secondary effect, which is a consequence of the existence of all the masses in the universe which create the intensity $I_o$. The primary effect, however, is the direct interaction between the masses $m_1$ and $m_2$ which results from the absorption of the GR emitted by $m_1$ and absorbed by $m_2$ (and vice versa) and leads to a repulsive force between $m_1$ and $m_2$. In this paper this effect is of interest for a special geometry of $m_1$ and $m_2$: The mass $m_1$ may exist with spherical symmetry inside a sphere with radius R. At the distance R from the centre, that means at the periphery of $m_1$, this mass $m_1$ produces a certain intensity of GR with a directional component $I_d$,

$$I_d = E * m_1 / R^2 \qquad (5)$$

as explained in the foregoing paper (Albers 1997). This directional GR from $m_1$ is absorbed by the mass $m_2$ which is assumed to be located at the periphery of $m_1$. Then the primary interaction between $m_1$ and $m_2$ is a repulsive force $F_r$ with

$$F_r = (E*m_1/R^2) * A*m_2 * P \qquad (6)$$

where the factor $(E*m_1/R^2)$ is identical with $I_d$ in equation (5).

The secondary, attractive force between $m_1$ and $m_2$ according to equation (3) and the primary, repulsive force according to equation (6) are both proportional to the product $m_1*m_2$. If for masses $m_1$ and $m_2$ sufficiently small compared to the masses of the universe the repulsive force according to equation (6) is smaller than the attractive force according to equation (3), this must not be valid if the mass $m_1$ is increased more and more. Somewhere the region of linearity of both forces with the product $m_1*m_2$ must be leaved. This becomes clear if one considers the case that the mass $m_1$ grows to an amount comparable to or larger than the mass of the universe. Then the repulsive interaction between $m_1$ and $m_2$ ( at their short distance R ) must exceed the attractive force between them.

The attractive forces produced according to equations (2) and (3) by all the masses $m_j$ of the universe can lead to mass accumulations. The primary, repulsive force discussed in connection with equation (6), however, gives some limit concerning mass and concentration of such mass-accumulations. A mass accumulation with mass $m_1$, radius R, and spherical symmetry is stable against the incorporation of a mass $m_2$ located at its periphery, and thus against an increase of the mass $M = m_1$ and the mean density if there is a balance between the attractive forces according to equation (3) and the repulsive forces according to equation (6). In equation (3) the term $I_0*A*m_1$ represents in this linearized form the part of Io which is absorbed by the mass $m_1$. If this factor is desribed by the correct exponential function of absorption it becomes evident that for increasing values of $m_1$ this factor saturates towards the limit $I_0$ which gives the upper bound for the attractive forces between $m_1$ and $m_2$. The simplest law concerning the balance between the attractive forces determined by $I_0$ and the repulsive forces produced by $I_d$ results from the assumption that the value of $I_d$ amounts to a certain part alpha of $I_0$:

$$I_d = alpha * I_o \qquad (7)$$

With r in equation (3) equal to R in equations (5) and (6), and $I_o$ equal to a constant in a first approximation, this leads to the result

$$M / R^2 = constant \qquad (8)$$

for possible mass accumulations in the universe with mass M and radius R. The spherical symmetry of such mass accumulations is a consequence of the fact that on large scales the mass distribution in the universe is rather homogeneous and thus of radial symmetry. Therefore also the attractive force according to equations (2) and (3) has spherical symmetry. Such mass accumulations are in equilibrium and therefore stable for ever if the mass distribution in the universe does not change. The mass M of such spherical systems is not limited to a narrow range of values but may possess very small or very high values depending on the amount of mass which existed anywhere near the condensation points of these systems.

## 2. SPHERICAL OBJECTS IN THE UNIVERSE

In the universe there exist several systems of mass accumulations with spherical symmetry, but quite different masses: Planets, stars, globular star clusters, elliptical galaxies, the almost elliptical cores of spiral galaxies, and the spherical clusters of galaxies. There exist huge differences concerning the density of these systems: The planets and stars possess mass-density values of the order of 1000 kg/m$^3$, whereas the density of the other systems extends from about $5*10^{-24}$ kg/m$^3$ for the largest clusters of galaxies to about $4*10^{-18}$ kg/m$^3$ for the smallest globular star clusters as can be calculated especially from the data shown in the last figure of this publication. These two values are much higher than the mean density of the universe with the above mentioned value of about $5*10^{-28}$ kg/m$^3$, but they are more than 20 orders of magnitude lower than the corresponding values of planets and stars. As already discussed in chapter 1 the efficiency to emit gravitational radiation may depend in an unkwown way on parameters as density, temperature, etc. The validity of relation (8) can only be examined for systems which are similar with respect to these parameters. In the following only the four last mentioned systems are analysed, the two first mentioned systems, planets and stars, however, are excluded because of the huge gap of more than 20 orders of magnitude between the values of the mean density of these two groups from the values of the other four groups of mass accumulations.

When considering the four largest groups mentioned above, globular star clusters, elliptical galaxies, the almost elliptical cores of spiral galaxies, and the spherical clusters of galaxies, one may recognize that especially all these systems contain old stars of population II, which may be as old as the universe itself. All these systems seem to exist in a state of equilibrium since long times, in agreement with the basic ideas of this paper.

The mass distribution of spiral galaxies with nonspherical symmetry, with old stars of population II in the core and young stars of population I in the spiral arms, is too complicated to be included into the following considerations. Therefore there remain three groups of mass accumulations which are of interest in the following: The globular star clusters, the elliptical galaxies, and the spherical clusters of galaxies. All three systems show the typical increase of the density towards their center in accordance with the alternative ideas in the foregoing paper (Albers 1997). There are some more similarities between these systems which are reflected in corresponding statements: The spherical clusters of galaxies are similar to the globular star clusters (Cambridge Enzyklopädie 1989); the shape of the elliptical galaxy M87 supports the assumption that it is a huge globular star cluster (same reference); the regular clusters of galaxies show a density law similar to that of elliptical galaxies (Roos, N. & Norman, C.A. 1979). Thus is seems not unrealistic that these three systems may be treated by one and the same model with its final conclusion represented by equation (8). The three systems are of very different size, where the smallest one, the globular star clusters, are typical components of elliptical galaxies and the elliptical galaxies themselves are the most common members in the largest systems, the spherical clusters of galaxies. The hierarchical arrangement of these spherical systems is not surprising because it reflects the possibility to build up stable systems of every size at every place with sufficiently high amount and density of matter.

For the examination of these three systems on the basis of equation (8) two values are needed, the radius R of the system and its mass M. In data compilations usually the apparent diameters D are specified and their original values can be used as well as the radius itself. The data may have, however, rather large limits of error which result from the necessary observations. The first error is due to uncertainties of the angular diameter because its determination may be influenced by the visual magnitude of the system or by the not well known galactic extinction. Second the distance between the system under consideration and our solar system is needed to calculate the absolute value of the diameter in parsec. And the different methods to calibrate distances include errors, the most commonly discussed one during the last years is the value of Hubble´s constant that is fundamental for the determination of the large distances to elliptical galaxies and to the spherical clusters of galaxies. These limits of error are mentioned primarily in order to indicate that due to the influence of the radius R on the results of equation (8) no accuracies in the region of percents can be expected.

The main error, however, must be expected in the second factor in equation (8), namely the mass M, and this error is of a systematical type. The masses of all three considered systems are usually determined by the application of the virial theorem on rotation curves or on the velocity dispersion of particles in such systems. The elliptical systems, however, have no pronounced rotational velocities as in the case of spiral galaxies where the values of the rotational velocities are the basis for the determination of the radius dependent mass. The crucial point in connection with every application of the virial theorem is the usual assumption of a universal gravitational constant G which is valid with the same value everywhere in the universe.

The alternative reflections on gravitation (Albers 1997) let expect that the universal gravitational constant G, if it is used in connection with Newton`s Law of Gravitation, must be replaced by a gravitational factor which is strongly dependent on the mass distribution particularly right next to the point of consideration. Especially near the center of spherical clusters of galaxies with their huge mass accumulations high values of the gravitational factor must be expected according to equation (1). Thus it is not surprising that for elliptical galaxies located in such regions extremely high mass-to-luminosity ratios M/L up to about 100 ( all mass-to-luminosity values in this paper are given in solar units ) are calculated by the conventional use of the virial theorem (Voigt 1982). But even higher values of M/L of about 330 ( adopting a realistic value of 75 km/s/Mpc of Hubble´s constant ) are reported from recent measurements of the mass to light ratio of the whole Abell 2218 system which is one of the largest spherical clusters of galaxies (Squires et al. 1996). If on the other hand mass determinations on the basis of gravitational effects are performed in regions with no remarkable increase of the mean density, for example in the outer part of our galaxy, quite normal values are obtained, e.g. for the globular star cluster M22 a value M/L of about 1 applies best for the cluster as a whole (Peterson & Cudworth 1993).

These so determined masses of the three different systems are really not suited for an inspection of equation (8). Instead of these masses the absolute visual magnitude may be a better parameter, especially if one reconsiders that all three systems are composed of old stars of population II with probably not very large differences of their mass-to-luminosity ratios. Therefore the inspection of equation (8) will be replaced by the inspection of the equation

$$L / D^2 = \text{constant} \qquad (9)$$

with the absolute luminosity L in units of the luminosity of the sun and the diameter D in parsec. At the end of this evaluation the data of all three systems will be changed from luminosity to mass values using a mass-to-luminosity ratio 1. This value was already assumed by W. Seggewiss as a realistic value for globular star clusters in the Landolt-Börnstein data compilation (Seggewiss 1982). The value itself is of no great importance, of more interest is the fact that one and the same value will be used for all three systems.

### 3. GLOBULAR STAR CLUSTERS

Globular star clusters are very common systems in the universe. In our own galaxy, the milky way, more than 100 of them exist which belong to the oldest objects in the universe. Their ages with estimated values near 15 Gyr (VandenBerg, Bolte, & Stetson 1996) are as high as the age of the universe determined from the value of Hubble´s constant. But also younger systems may exist according to the own alternative reflections on gravitation. And such systems really exist; for example in the Magellanic Clouds globular clusters with ages of only $10^7$ to $10^8$ yr are observed (Unsöld & Baschek 1991). There exist also much larger groups of globular star clusters than that in our galaxy. About 1800 candidates of the globular cluster system of the huge elliptical galaxy NGC 4472 ( M49 ) could be investigated in detail. And this number seems to be only 1/4 of the total number of globular clusters in this galaxy (Geisler, Lee, & Kim 1996). In the following of course only the globular clusters in our galaxy are of interest because only for them the necessary data are known well enough.

In the Encyclopedia of Physics (1959) data of 118 globular clusters are presented by Sawyer Hogg on the basis of her earlier compilation (Sawyer 1947). A list with data of only 70 globular clusters can be found in the Landolt-Börnstein data compilation (Landolt-Börnstein 1965). In this paper, however, a more recent data compilation will be used which contains carefully revised data and is available through the Internet. It can be found on the commercially available CD-ROM GUIDE 4.0 and contains the detailed data presented by Ruprecht, Balazs and White (1981) together with an extended list of references. The CD-ROM contains a machine-readable version of a 1437 kByte file (Globular.dat) from which the relevant data were read by a basic-program. In order to get the most reliable data all clusters were left out of consideration for which no integral spectral type is reported in connection with the citation Sawyer Hogg and the year 1963. In this way from the list of 137 clusters only 64 remain for which the corresponding data exist. The globular cluster GCL-46 is omitted because no value of the linear diameter is given, GCL-87 is neglected because of the missing data of the integrated magnitude. So there remain 62 clusters for with the necessary values are available. The linear diameter LD in parsec in the compilation is taken as diameter D in equation ( 9). From the integrated magnitude MT and the distance DIST in kpc in the compilation the luminosity L is calculated according to

$$L / L_{Sun} = \exp( 0.4*( m_{Sun} - MT + 10 + 5*\log(DIST))). \qquad (10)$$

Here $L_{Sun}$ and $m_{Sun}$ = 4.87 mag are the luminosity and the absolute magnitude of the sun, respectively.

In Figure 1 the values of the luminosity L of the above mentioned 62 globular star clusters are drawn against the linear diameter D, both in logarithmic scales. The straight line is no best fit to the data, but it serves as a guide to the eye and its increase corresponds exactly to the power 2 of the dependence of L on D derived in equation (9). The scatter of the data is rather high, but it seems that the data are not incompatible with the statement expressed by equation (9).

## 4. ELLIPTICAL GALAXIES

A suitable compilation of galaxies can be obtained through the Internet from the Astronomical Data Center ( ADC 1996). It is the electronic version of the Third Reference Catalogue of Bright Galaxies ( RC3 ) which was compiled by de Vaucouleurs et al. ( 1991 ) and published in printed form by the Springer-Verlag. It contains more than 23 thousand entries from which easily those elliptical galaxies can be extracted which are of type E0 and thus of exactly spherical form. Of course also slightly distorted elliptical galaxies of type E1 etc. could be taken into consideration because their deviation from the exactly spherical shape is smaller than the probable uncertainty of other data of the E0-galaxies. The limitation to the E0-type, however, restricts the selection to the really spherical systems discussed in equations (8) and (9) and it reduces the number of galaxies to a value that can be shown graphically. The compilation consists of the two files Rc39a and Rc39b with ascending values of right ascension. The search for the expression „E.0" in the column „type" delivers 31 galaxies from Rc39a and 44 ones from Rc39b. Only the Rc39b data will be used in the following in order to get not too many records.

From the V3K-value of the escape velocity the distance d is calculated using Hubble´s constant H with a value of 75 km/s/Mpc according to

$$d = V3K / H. \qquad (10)$$

From the log D25 value of the angular diameter the absolute diameter in pc follows by use of the distance d. The B value of the photographic magnitude, or if missing, the total B magnitude BT are used together with the distance d to calculate the absolute magnitude of the galaxy and its luminosity in solar units. For 8 galaxies the necessary data about magnitude or V3K are missing so that there remain 36 galaxies, the data of which are shown in Figure 2.

In spite of the rather large scatter of the values it may be stated that also these data seem to be compatible with a quadratic power law, though it should be noted that most points are lying slightly below the straight line which is a prolongation of the straight line in Figure 1 and which represents a quadratic dependence of the luminosity L on the diameter D.

## 5. SPHERICAL CLUSTERS OF GALAXIES

The two best known compilations of clusters of galaxies are those originally compiled by Abell and by Zwicky which can be obtained in a modern form with revised data through the Internet (ADC 1996). Though the number of clusters is larger in the Zwicky catalogue, the Abell catalogue of rich, compact clusters of galaxies is the right one for this paper because the rich clusters concentrate especially on the regular systems with spherical symmetry. Although the data which were extracted from the Abell catalogue seemed to be suitable to support the ideas of this paper, some other data will be used here in order to acknowledge the observations and conclusions which were performed already by E.F. Carpenter almost 60 years ago and which are described in the Cambridge Enzyklopädie (1989). Not all the statements of Carpenter cited in this encyclopadia which are conform with the ideas in this paper can be reproduced here. Besides Carpenter´s statement that there probably exists a sharp upper boundary of the number of galaxies in dependence on the diameter of the system only his data which are shown graphically in the above mentioned encyclopedia will be considered here. According to these data, the number N of galaxies in a cluster is drawn against the diameter D of the cluster in Figure 3. The straight line inserted into this figure serves as a guide to the eye and corresponds to a quadratic dependence of N on D. The relatively large scatter of the data prevents the selection of a special functional dependence of N on D by least squares methods. But it can also be stated that a quadratic dependence of N on D, as represented by the straight line in Figure 3, is a rather good fit to the data, especially if not to much significance is attached to the number N in systems with N < 10 for which the definition of a diameter D is obviously not so easy.

There exists no direct correlation between the ordinates in Figures 1 and 2 (luminosity L) and the ordinate in Figure 3, the number N of galaxies. Such a correlation, however, is necessary to combine the data of Figures 1 and 2 with the data of Figure 3. Elliptical galaxies, including those of the E0-type in Figure 2, are the predominating components in clusters of galaxies. Therefore it seems reasonable for a first approach to correlate the luminosity L of a cluster of galaxies in Figure 3 to the luminosity of the E0-type galaxies in Figure 2. The mean arithmetical value of the luminosity L of all 36 elliptical galaxies in Figure 2 is L = $1.43*10^{10}L_{Sun}$. Assuming this value as a mean value of the N galaxies of every particular cluster in Figure 3 one gets the luminosity values of L of the ordinate at the right side in Figure 3.

## 6. COLLECTION OF SPHERICAL OBJECTS

The three different systems of spherical mass accumulation characterized by their data in the Figures 1 to 3 are derived from the same basic assumptions represented by equations (8) and (9). Therefore it should be possible to merge the data of all three systems into one figure in which the quadratic dependence of the luminosity L on the diameter D is described by the same value of the constant in equation (9). In Figure 4 all the data from Figures 1 to 3 are shown in one drawing together with a straight line which is identical with the lines in the foregoing pictures. As can be seen from Figure 4 this line with an exactly quadratic increase of the luminosity L on the diameter D is well suited to describe all three sytems in the same way in agreement with the expectations described by equation (9). Due to the enormous range in diameter and luminosity, in this figure the deviations of the single points from the straight line are by far not so impressive as in the foregoing figures.

There are several causes of the large scatter of the data around the predicted straight line. First it must be considered that the determination of data of luminosities and diameters are strongly influenced by corrections due to the galactic extinction. This correction with values up to more than 3 mag, e. g. for the data of globular clusters (Encyclopedia of Physics 1959), is not well known but it can influence the luminosity values by a factor of more than ten. A second point which should be mentioned here is the uncertainty about the determination of the distance, especially if the value of Hubble´s constant has a decisive influence on its value. A third point of interest in connection with Figure 4 is the fact that the value on the right side of equation (9) is a constant only in a first approximation. In a second approximation it should be considered that this constant is proportional to $I_o$ and thus depending especially on the mass distribution existing near the object of consideration. Therefore it is not surprising that the highest values of $L/D^2$ in Figure 1 exist for those globular star clusters which are located close to the galactic centre, with a shortest distance of only 1.3 kpc ( here a distance of 8.5 kpc is assumed as the distance between our solar system and the galactic centre). At such a position, already inside the central bulge of the milky way - with a radius of about 2.5 kpc -, the value of $I_o$ can exceed remarkably that one valid for other globular star clusters located at positions more distant from the galactic centre.

For the second group of objects, the elliptical galaxies of EO-type, similar systematic corrections of the data in Figures 2 and 4 seem to be appropriate. The two highest values of $L/D^2$ in Figure 2 are that of PGC 39188 and PGC 43514, the two lowest ones belong to PGC 62122 and PGC 63852, with a factor of about ten between the highest and the lowest values. If

12one takes a look at the drawings in URANOMETRIA 2000.0 or if one displays a region with a diameter of 2 deg with the program GUIDE on the CD-ROM GUIDE 4.0 one can detect the possible cause of this high factor: The diagrams centered at the first two galaxies show huge accumulations of galaxies and clusters of galaxies whereas around the latter two galaxies such mass accumulations are missing.

Alltogether these three points should serve here only to show that by use of all the experimental data presented in Figure 4 not more than a very rough correspondence with the statements and equations of the alternative approach in this paper can be expected. Careful corrections corresponding to these three or even more points are necessary to decide how well the experimental data are in agreement with the postulated statements. But that is far beyond the scope of this paper.

A further uncertainty occurs when moving on from the luminosity data to mass data of the objects in Figure 4. Globular star clusters and elliptical galaxies are usually characterized by their spectral class but from this value no direct calculation of their masses is possible as in the case of stars by use of the Hertzsprung-Russel diagram. And the determination of the overall mass M on the basis of the different branches of stars in globular star clusters and in elliptical galaxies is far to awkward to be performed here. Therefore a scale for the mass M is added to Figure 4 using a M/L ratio equal to that of our sun. As mentioned above this value M/L = 1 in solar units was already assumed to be appropriate for globular star clusters in the Landolt-Börnstein data compilation (Seggewiss 1982). And this value is adopted here as a realistic one also for the other two groups of objects in Figure 4. In this way the mass-diameter relation mentioned in the title is obtained with the expected final result that for the different spherical systems in the universe the mass is quadratically dependent on the diameter of the system.

## 7. CONCLUSIONS

On the basis of the alternative reflections on gravitation developed in the foregoing paper (Albers 1997) it is assumed that stable spherical mass accumulations may exist in the universe. Indeed such objects can be observed in an incredible high number in the universe in the form of globular star clusters, elliptical galaxies and spherical clusters of galaxies. The most simple conclusion concerning the stability of these objects against mass-density changes by competitive attractive and repulsive forces leads to a quadratic dependence of the masses of these objects on their diameter. It is shown in this paper that the corresponding experimental data of these objects are in agreement with this conclusion within the rather high limit of error resulting from different sources of uncertainty. It can be noted here, however, that neither the existence nor any parameter of these really existent stable objects can be derived on the basis of Newtonian mechanics without the assistance of additional assumptions. A prominent example is the process of heating by binary stars to explain the stability of the stellar density inside globular clusters (Guhathakurta et al.1996).

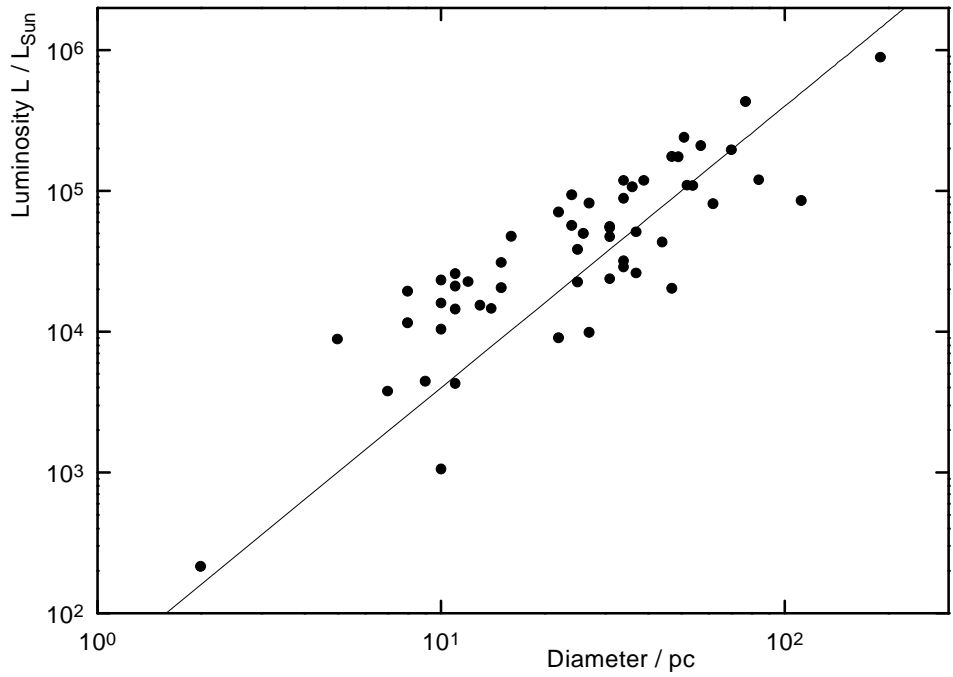

FIG. 1.- Luminosity L ( in solar units $L_{sun}$ ) of 62 globular star clusters drawn against the linear diameter D ( in parsec ) of the cluster. Straight line: Guide to the eye representing a quadratic dependence of L on D ( no best fit ).

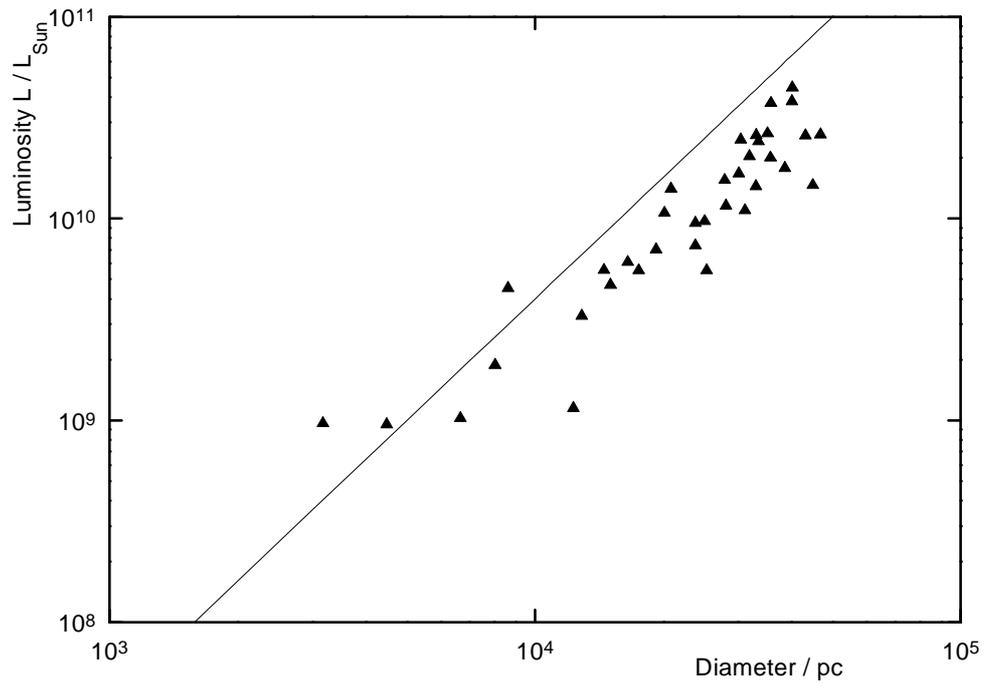

FIG. 2.- Luminosity L ( in solar units $L_{sun}$ ) of 36 elliptical galaxies of the E0-type drawn against their linear diameter D ( in parsec ). Straight line: prolongation of the straight line in FIG.1 with the quadratic dependence of L on D ( no best fit to these data).

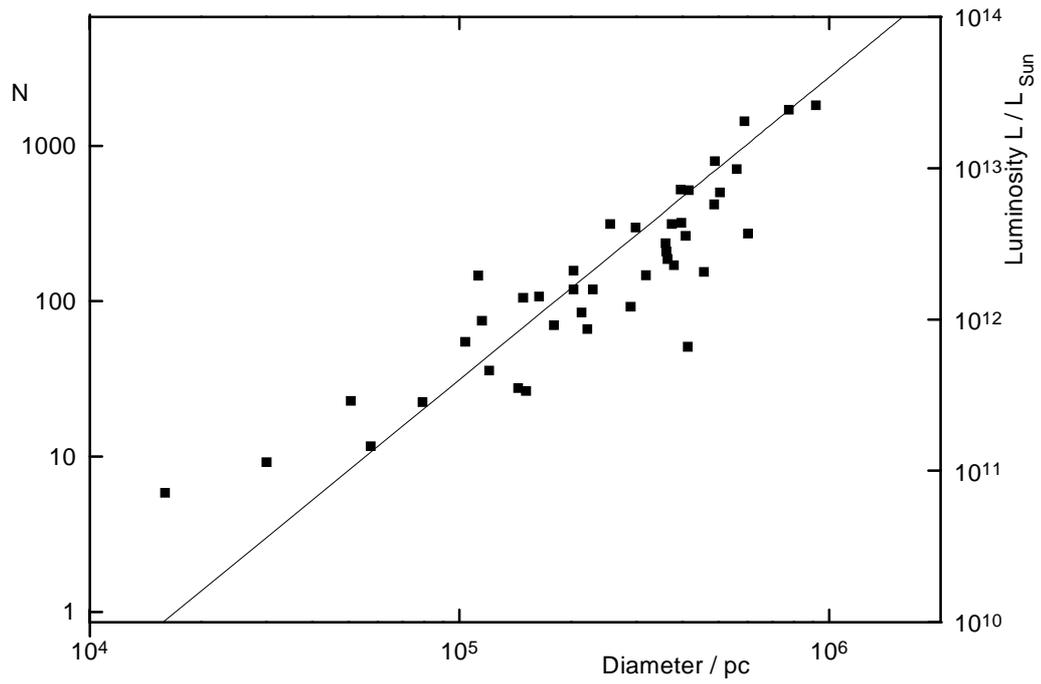

FIG. 3.- Number N of galaxies in clusters of galaxies drawn against the diameter D ( in parsec ) of the cluster. The straight line corresponds to a quadratic dependence of L on D ( no best fit to these data). Data from Cambridge Enzyklopädie (1989). The values of the luminosity L at the right ordinate are based on the mean value of L in FIG. 2.

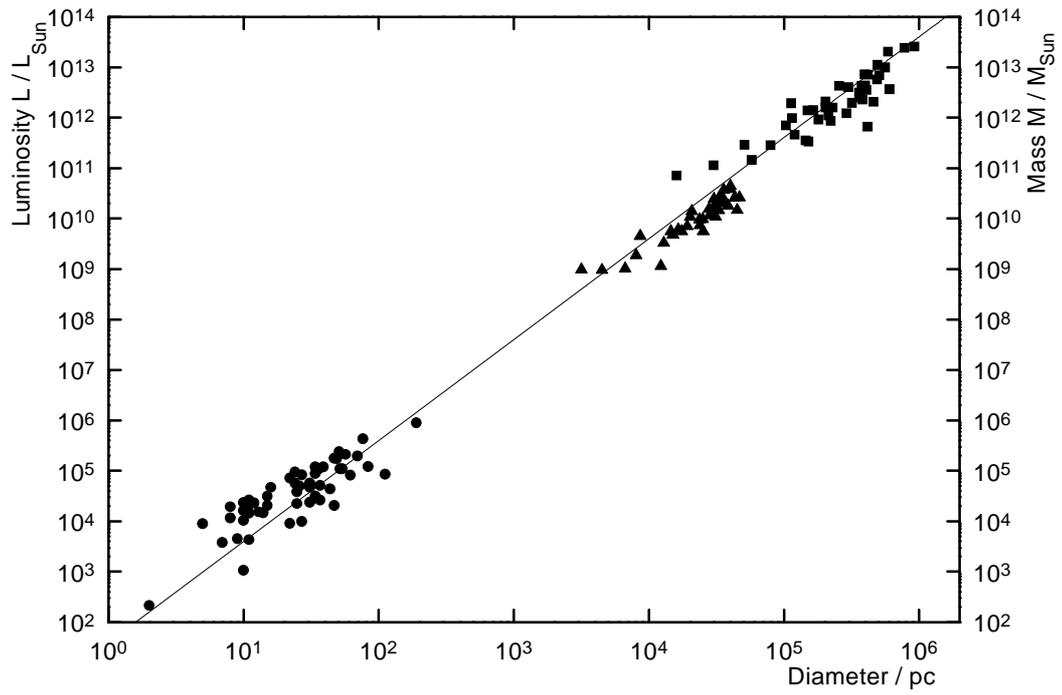

FIG. 4.- Dependence of the luminosity L and the mass M on the diameter D of three groups of spherical objects in the universe. Circles: globular star clusters, triangles: elliptical galaxies of E0-type, squares: spherical clusters of galaxies. The mass values M at the right ordinate result from the L-values at the left ordinate by M/L = 1 in solar units.